\documentclass{PoS}

\usepackage{graphicx}
\usepackage{subfigure}

\title{The lattice gluon propagator in Landau gauge at zero and finite temperature}

\ShortTitle{The lattice gluon propagator in Landau gauge at zero and finite temperature}

\author{\speaker{Paulo J. Silva}\\
CFC, Departamento de F\'{\i}sica, Faculdade de Ci\^encias e Tecnologia, Universidade de Coimbra, 3004-516 Coimbra, Portugal\\        
        E-mail: \email{psilva@teor.fis.uc.pt}}

\author{Orlando Oliveira \\%
CFC, Departamento de F\'{\i}sica, Faculdade de Ci\^encias e Tecnologia, Universidade de Coimbra, 3004-516 Coimbra, Portugal\\
        E-mail: \email{orlando@teor.fis.uc.pt}}

\abstract{The interplay between the finite volume and finite lattice spacing is investigated using lattice QCD simulations to compute the Landau gauge gluon propagator at zero temperature. Comparing several ensembles with different lattice spacings and physical volumes, we conclude that the dominant effects, in the infrared region, are associated with the use of a finite lattice spacing. The simulations show that decreasing the lattice spacing, while keeping the same physical volume, leads to an enhancement of the infrared gluon propagator. Moreover, we also present results for the Landau gauge gluon propagator at finite temperature.}

\FullConference{Xth Quark Confinement and the Hadron Spectrum\\
		 8-12 October 2012\\
		 TUM Campus Garching, Munich, Germany}

\begin{document}

\section{The gluon propagator at zero temperature}

In \cite{OliSi12}, we have studied the lattice Landau gauge gluon propagator, using several ensembles with different lattice volumes and lattice spacings. Here we will report on the comparison of the gluon propagator computed from similar physical volumes but different lattice spacings. For details on the lattice setup see  \cite{OliSi12}.

\begin{figure}[b]
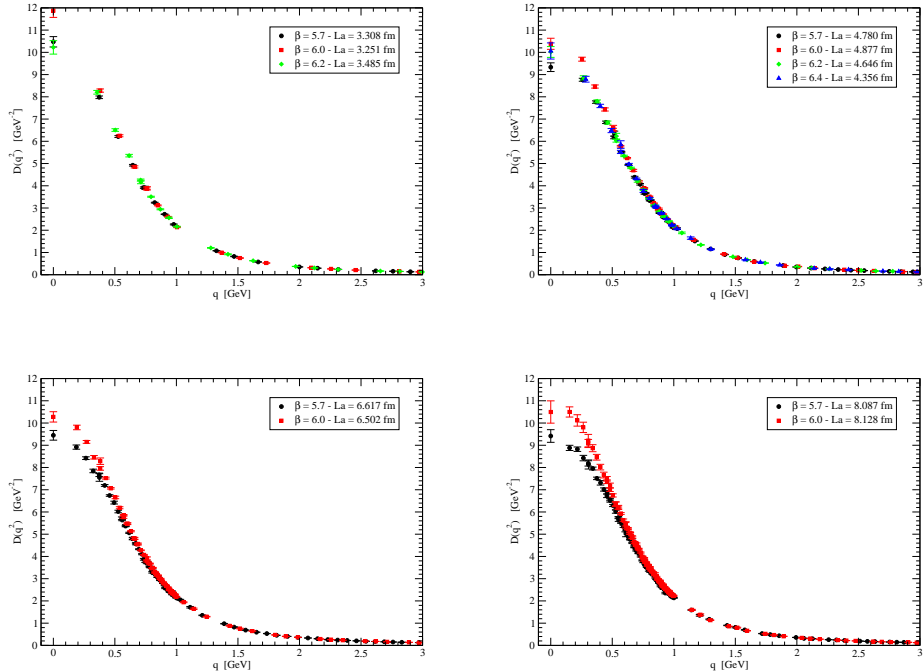
 
   \centering
   \subfigure{ \includegraphics[scale=0.235]{figures/glue_R4GeV_V3.3fm.eps} } \qquad
   \subfigure{ \includegraphics[scale=0.235]{figures/glue_R4GeV_V4.6fm.eps} }

\vspace{0.6cm}
   \subfigure{ \includegraphics[scale=0.235]{figures/glue_R4GeV_V6.6fm.eps} } \qquad
   \subfigure{ \includegraphics[scale=0.235]{figures/glue_R4GeV_V8.1fm.eps} }
  \caption{Comparing the renormalized gluon propagator at $\mu = 4$ GeV for various lattice spacings and similar physical volumes.}
   \label{fig:gluespac}
\end{figure}

Figure \ref{fig:gluespac} shows the gluon propagator, renormalized at 4 GeV,  for physical volumes ranging from 3.3 fm to 8.1 fm. In all cases, for momenta above $\sim900$ MeV the lattice data define a unique curve; this means that the renormalization procedure has been able to remove all dependence on the ultraviolet cutoff for the mid and high momentum regions.

However, we observe in the infrared region sizeable effects due to the finite lattice spacing of the simulation. In particular, the plots show that large lattice spacing simulations underestimate the propagator in the infrared region. Moreover, we found that the corrections due to the finite lattice spacing are larger than the corrections due to the finite volume of the lattice simulation --- for details see \cite{OliSi12}.

One can ask whether the analysis reported here depends on the renormalization scale. Usually, one chooses a renormalization point $\mu$ in the high momentum region, in order to allow a contact with the perturbative behaviour. However, choosing a renormalization point in the infrared is also possible. Here we consider two alternative renormalization points,  $\mu = 500$ MeV and
$\mu = 1$ GeV and study how the propagators differ. Note that we only show results for the simulations performed at $La \approx 8$ fm.

\begin{figure}[t]
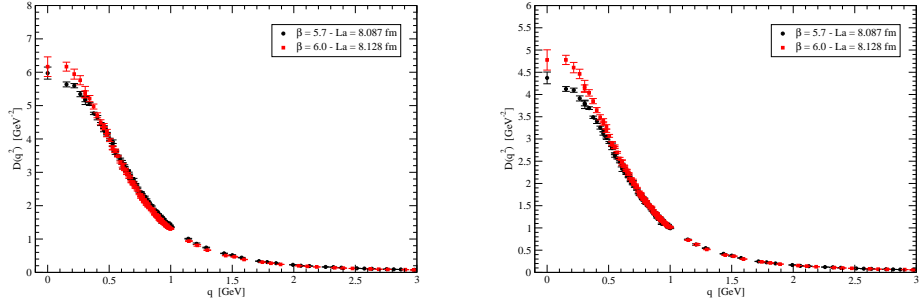
 
   \centering
   \subfigure{ \includegraphics[scale=0.235]{figures/glue_R500MeV_8fm.eps} } \qquad
   \subfigure{ \includegraphics[scale=0.235]{figures/glue_R1GeV_8fm.eps} }
  \caption{Gluon propagator renormalised at $\mu = 500$ MeV (left) and at $\mu = 1$ GeV  (right).}
   \label{fig:glue1GeV500MeV}
\end{figure}

\begin{figure}[b]
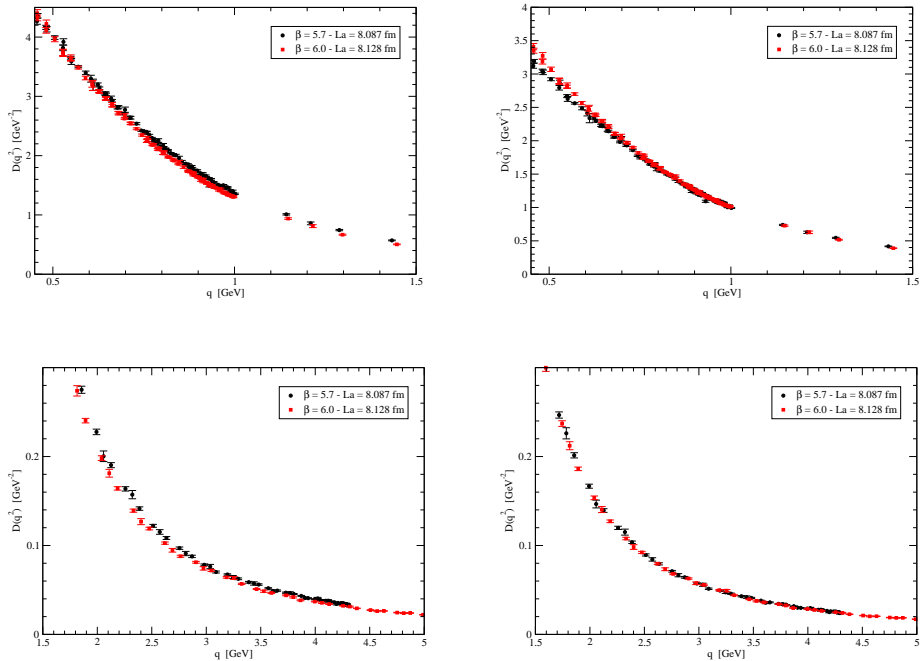
 
   \centering
   \subfigure{ \includegraphics[scale=0.235]{figures/glue_R500MeV_8fm_ZOOM.eps} }\qquad
   \subfigure{ \includegraphics[scale=0.235]{figures/glue_R1GeV_8fm_ZOOM.eps}  }
   
   \vspace{0.5cm}
   \subfigure{ \includegraphics[scale=0.235]{figures/glue_R500MeV_8fm_ZOOM_UV.eps} }\qquad
   \subfigure{ \includegraphics[scale=0.235]{figures/glue_R1GeV_8fm_ZOOM_UV.eps}  }
  \caption{The gluon propagator renormalised at $\mu = 500$ MeV (left) and at $\mu = 1$ GeV  (right) 
                at intermediate momentum range (top) and up to 5 GeV (bottom).}
   \label{fig:glue1GeV500MeV_IR}
\end{figure}

Figures \ref{fig:glue1GeV500MeV} and \ref{fig:glue1GeV500MeV_IR} show that for both renormalization points, the differences in the infrared region are well beyond one standard deviation. This means that the dominant effect in the infrared propagator is indeed due to the finite lattice spacing, and not the physical box size of the simulation; this conclusion is independent of the renormalization scale. Therefore lattice artifacts due to a finite lattice spacing play an important role in the determination of the infrared behaviour of the gluon propagator in the Landau gauge. 

\clearpage

\section{The gluon propagator at finite temperature}

\begin{table}[b]
\begin{center}
\begin{tabular}{cccccc}
\hline
Temp. (MeV) &	$\beta$ & $L_s$ &  $L_t$ & a [fm] & 1/a (GeV) \\
\hline
121 &	6.0000 & 32,64 & 	16 & 	0.1016 &  	1.9426 \\
162 &	6.0000 & 32,64 & 	12 & 	0.1016 & 	1.9426 \\
243 &	6.0000 & 32,64 & 	8 & 	0.1016 & 	1.9426 \\
260 &	6.0347 & 68    & 	8 & 	0.09502 & 	2.0767 \\
265 &	5.8876 & 52    & 	6 & 	0.1243 & 	1.5881 \\
275 &	6.0684 & 72    & 	8 & 	0.08974 & 	2.1989 \\
285 &	5.9266 & 56    & 	6 & 	0.1154 & 	1.7103 \\
290 &	6.1009 & 76    & 	8 & 	0.08502 & 	2.3211 \\
305 &	5.9640 & 60    & 	6 & 	0.1077	 &      1.8324 \\
305 &	6.1326 & 80    & 	8 & 	0.08077 & 	2.4432 \\
324 &	6.0000 & 32,64 & 	6 & 	0.1016	 &      1.9426 \\
486 &	6.0000 & 32,64 & 	4 & 	0.1016	 &      1.9426 \\
\hline
\end{tabular}
\end{center}
\label{tempsetup}
\caption{Lattice setup used for the computation of the gluon propagator at finite temperature.}
\end{table}

Besides the Landau gauge gluon propagator at zero temperature, we are also engaged in the computation of the gluon propagator at finite temperature, in order to understand how temperature changes the gluon propagator \cite{eqcd12, latt12}. 

Finite temperature is introduced on the lattice through the reduction of the temporal lattice size: the simulations are done on lattices $L_s^3 \times L_t$, with $L_t \ll L_s$. The temperature is defined by $T=1/a L_t$.

At finite temperature, the gluon propagator is described by two tensor structures, 

\begin{equation}
D^{ab}_{\mu\nu}(q)=\delta^{ab}\left(P^{T}_{\mu\nu} D_{T}(q_4,\vec{q})+P^{L}_{\mu\nu} D_{L}(q_4,\vec{q}) \right) \nonumber
\label{tens-struct}
\end{equation}
where the transverse and longitudinal projectors are defined by
\begin{equation}
P^{T}_{\mu\nu} = (1-\delta_{\mu 4})(1-\delta_{\nu 4})\left(\delta_{\mu \nu}-\frac{q_\mu q_\nu}{\vec{q}^2}\right) \quad , \quad
P^{L}_{\mu\nu} = \left(\delta_{\mu \nu}-\frac{q_\mu q_\nu}{{q}^2}\right) - P^{T}_{\mu\nu} \, ;
\label{long-proj}
\end{equation}
the transverse $D_T$ and longitudinal  $D_L$ propagators are given by
\begin{equation}
D_T(q)=\frac{1}{2V(N_c^2-1)}\left(\langle A_i^a(q) A_i^a(-q)\rangle-\frac{q_4^2}{\vec{q}^2} \langle A_4^a(q) A_4^a(-q)\rangle \right) \nonumber
\end{equation}

\begin{equation}
D_L(q)=\frac{1}{V(N_c^2-1)}\left(1+\frac{q_4^2}{\vec{q}^2} \langle A_4^a(q) A_4^a(-q)\rangle\right) \nonumber
\end{equation}

 The lattice setup is shown in table \ref{tempsetup}. The finite temperature simulations described in this section have been performed with the help of Chroma library \cite{chroma}; the FFT transforms have been done with the PFFT library \cite{pfft}. For the determination of the lattice spacing we fit the string tension data in \cite{bali92}, using the functional form used in \cite{neccosommer}, in order to have a function $a(\beta)$; the fit has a $\chi^2/dof\sim0.03$. 

The reader should note that the simulation parameters have been carefully chosen, such that we only consider two different spatial physical volumes: $\sim(3.3\mbox{fm})^3$ and $\sim(6.5\mbox{fm})^3$. This allows for a better control of finite size effects.

\begin{figure}[t]
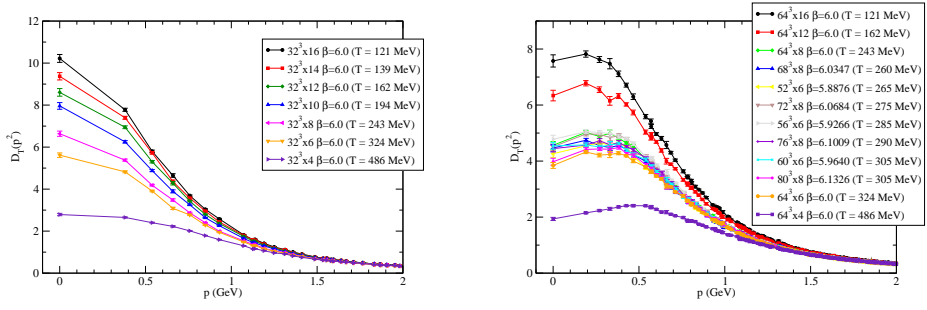
 
   \centering
   \subfigure{ \includegraphics[scale=0.222]{figtemp/trans32.eps} } \qquad
   \subfigure{ \includegraphics[scale=0.222]{figtemp/trans64.eps} }
  \caption{Transverse gluon propagator for  $\sim(3.3\mbox{fm})^3$ (left) and $\sim(6.5\mbox{fm})^3$ (right) spatial lattice volumes.}
   \label{fig:transtemp}
\end{figure}

\begin{figure}[htb]
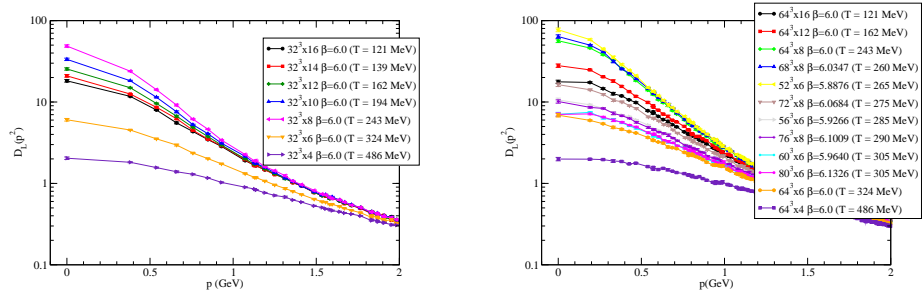
 
   \centering
   \subfigure{ \includegraphics[scale=0.215]{figtemp/long32.eps} } \qquad
   \subfigure{ \includegraphics[scale=0.215]{figtemp/long64.eps} }
  \caption{Longitudinal gluon propagator for  $\sim(3.3\mbox{fm})^3$ (left) and $\sim(6.5\mbox{fm})^3$ (right) spatial lattice volumes.}
   \label{fig:longtemp}
\end{figure}

\begin{figure}[htb] 
   \centering
   \subfigure{ \includegraphics[scale=0.25]{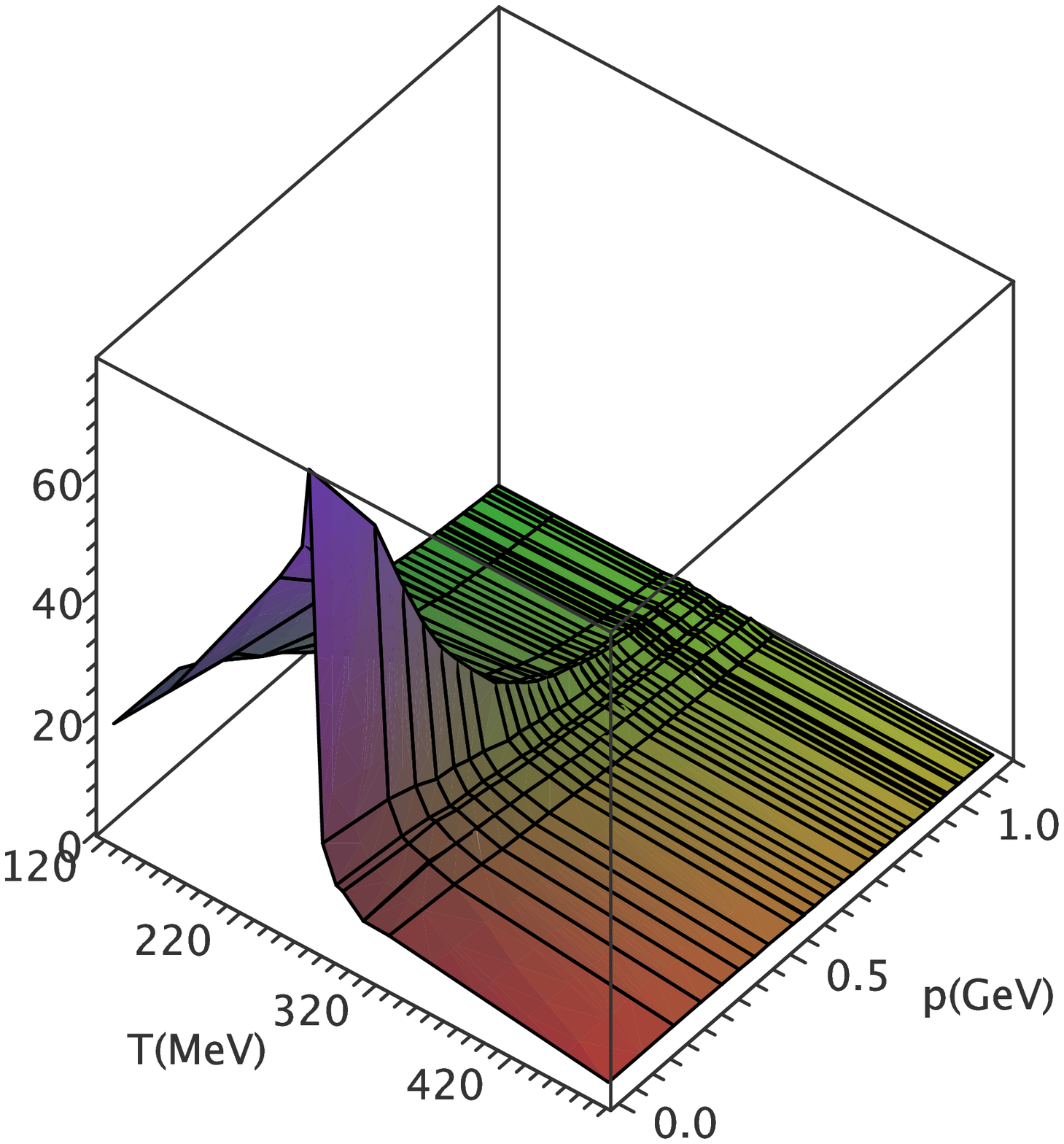} } \qquad
   \subfigure{ \includegraphics[scale=0.25]{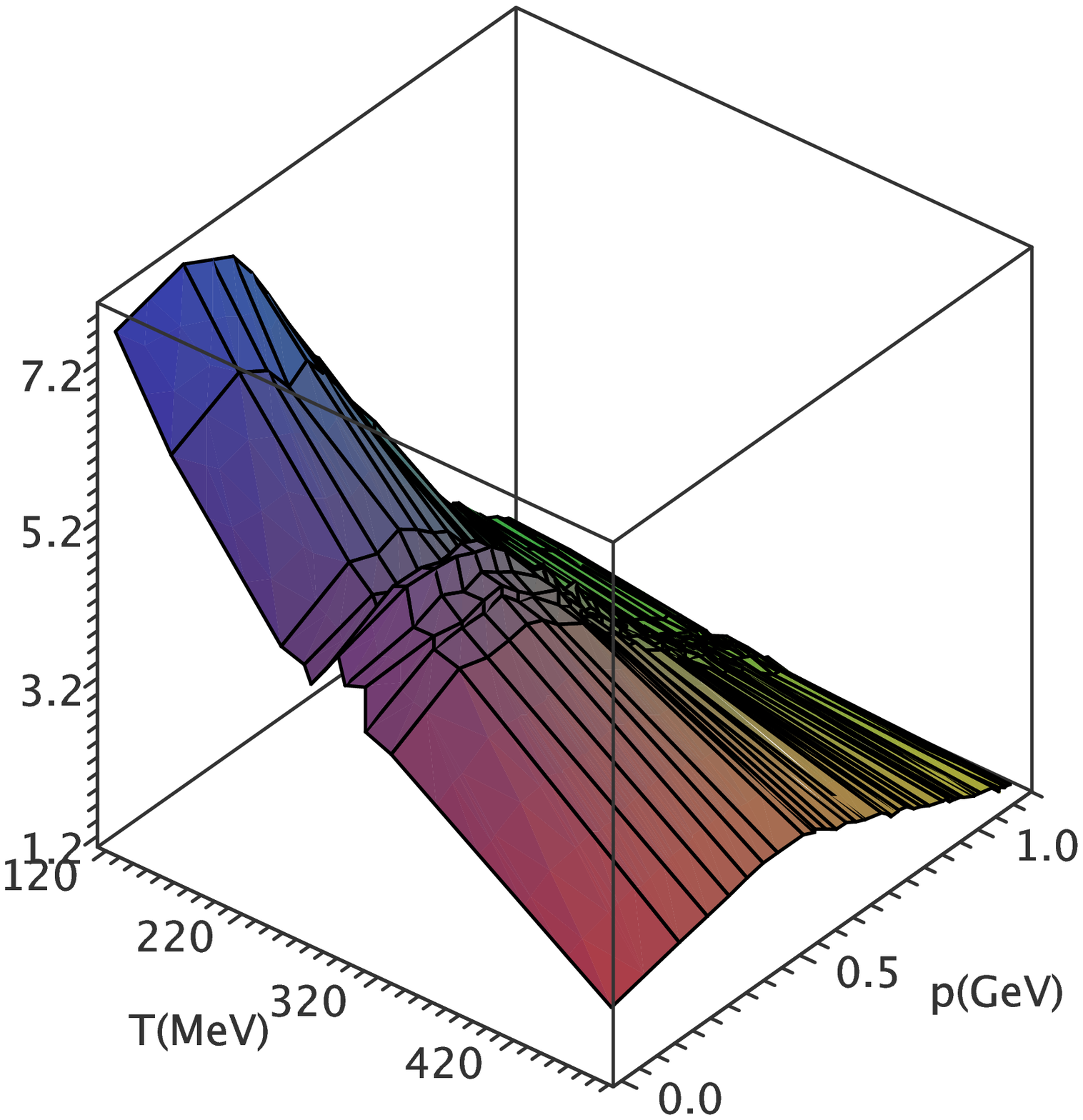} }
  \caption{Longitudinal (left) and transverse (right) gluon propagator as a function of momentum and temperature for a $\sim(6.5\mbox{fm})^3$ spatial lattice volume.}
   \label{fig:3dtemp}
\end{figure}

In figures \ref{fig:transtemp} and \ref{fig:longtemp} we show our results. In what concerns the transverse propagator, it decreases with the temperature for low momenta. Moreover, this component shows finite volume effects; in particular, the large volume data exhibits a turnover in the infrared, not seen at the small volume data. The longitudinal component increases for temperatures below $T_c\sim 270\, \mbox{MeV}$. Around $T_c$, there is a discontinuity and the propagator decreases for $T > T_c$. The behaviour of the gluon propagator as a function of the temperature can also be seen in the 3d plots shown in figure \ref{fig:3dtemp}.

\vspace{0.1cm}

\vspace{-0.1cm}

\begin{figure}[t]
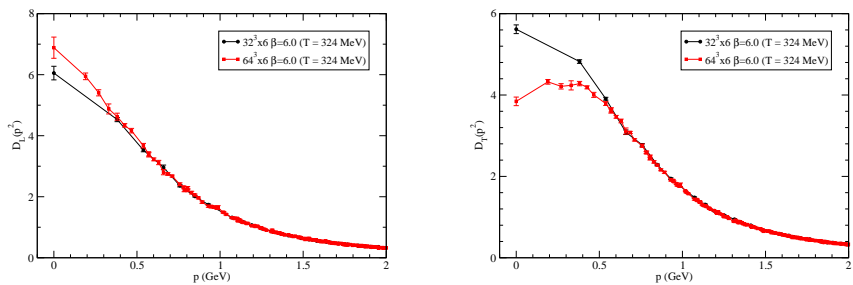
 
   \centering
   \subfigure{ \includegraphics[scale=0.215]{figtemp/long324MeV.eps} } \qquad
   \subfigure{ \includegraphics[scale=0.215]{figtemp/trans324MeV.eps} }
  \caption{Longitudinal (left) and transverse (right) gluon propagator for different spatial lattice volumes at T=324 MeV.}
   \label{fig:finvoltemp}
\end{figure}

The reported finite volume effects can be seen in more detail 
in figure \ref{fig:finvoltemp} for a temperature of 324 MeV. 

\begin{figure}[h]
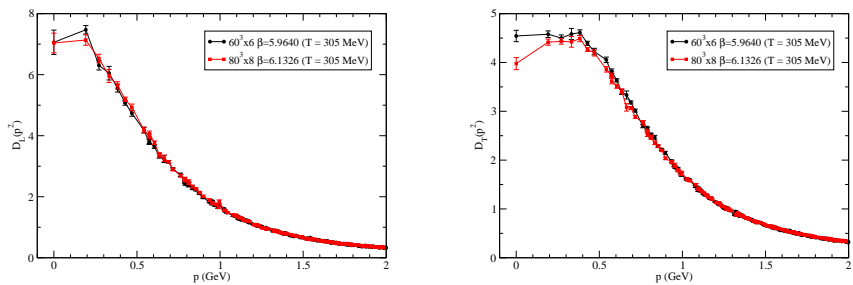
 
   \centering
   \subfigure{ \includegraphics[scale=0.215]{figtemp/long305MeV.eps} } \qquad
   \subfigure{ \includegraphics[scale=0.215]{figtemp/trans305MeV.eps} }
  \caption{Longitudinal (left) and transverse (right) gluon propagator for different lattice spacings (but similar physical volume) at T=305 MeV.}
   \label{fig:lattspactemp}
\end{figure}

A complete analysis should include a study of finite lattice spacing 
effects. For such a goal, we worked out two different simulations, 
at T=305 MeV, with similar physical volumes but different lattice spacings. 
Figure \ref{fig:lattspactemp} shows that, for this temperature, finite 
lattice spacing effects seem to be under control, with the exception 
of the zero momentum for the transverse component.

\vspace*{0.5cm}

Our preliminary results at finite temperature are rather similar to previous works --- see for example \cite{cucc} and references therein. We are currently working on a complete analysis of our data, towards a better understanding of lattice effects on the gluon propagator at finite temperature.

\section*{Acknowledgements}
 
Paulo Silva acknowledges support 
by FCT under contract SFRH/BPD/40998/2007. Work supported by projects 
CERN/FP/123612/2011, CERN/FP/123620/2011 and PTDC/FIS/100968/2008, 
projects developed under initiative QREN financed by UE/FEDER through 
Programme COMPETE.

\end{document}